\documentclass[aps,prl,twocolumn,reprint,showpacs,superscriptaddress]{revtex4-2}
\usepackage{amsmath,amssymb,graphicx,color}
\usepackage{times}
\usepackage{latexsym}
\usepackage{stmaryrd}
\usepackage[table]{xcolor}
\usepackage[colorlinks,allcolors=blue]{hyperref}
\usepackage{amsfonts}
\usepackage{natbib}
\usepackage{appendix}
\usepackage{dsfont}

\usepackage[T1]{fontenc}

\newcommand\ket[1]{\left|\textstyle{#1}\right\rangle}

\begin{document} 

\title{Emergent continuous symmetry and ground-state factorization induced by long-range interactions}

\author{Yue Yu}
\affiliation{Department of Engineering \& Applied Sciences, Yale University, New Haven, CT 06520, USA}
\affiliation{Division of Natural and Applied Sciences, Duke Kunshan University, Kunshan, Jiangsu, 215300 China}
\author{Myung-Joong Hwang}
\email{myungjoong.hwang@duke.edu}
\affiliation{Division of Natural and Applied Sciences, Duke Kunshan University, Kunshan, Jiangsu, 215300 China}

\date{\today}

\begin{abstract}
The spontaneous breaking of a $Z_2$ symmetry typically gives rise to emergent excitations possessing the same symmetry with a renormalized mass. Contrary to this conventional wisdom, we present a theory in which the low-lying excitation in the broken-symmetry phase acquires a continuous symmetry, even when the underlying symmetry of the system is discrete. In the presence of anisotropic long-range interactions, the order parameter renormalizes the relative strength of the particle-conserving and particle-nonconserving interactions. When one of the two renormalized interactions vanishes, a conservation law absent in the original Hamiltonian emerges, giving rise to a continuous symmetry. A striking consequence of the emergent continuous symmetry and conservation law is that it constrains quantum correlations in the ground-state to be zero, leading to the ground-state factorization in the presence of strong interactions. Our finding is a universal feature of quantum phase transitions in fully-connected systems and their lattice generalizations; therefore, it can be observed in a wide range of physical systems, including cavity QED systems and ion-traps.
\end{abstract}
\maketitle

{\it Introduction.---}Understanding the emergent phenomena of a system undergoing a phase transition is an important problem in physics. A key question in this context is: what are the properties of an order parameter and its fluctuations in broken symmetry phases. For a system possessing a continuous symmetry, the Goldstone theorem guaranties that a gapless excitation called the Goldstone mode~\cite{goldstone1961field,goldstone1962broken,leonard2017monitoring} always appears in the broken symmetry phase. For a discrete $Z_2$ symmetry case, this is no longer the case, as it is not possible to continuously deform from one possible state into another; instead, there typically exists a duality between the symmetric and broken symmetric phases, whereby the elementary excitation in the broken symmetry phase is described by the Hamiltonian with the same $Z_2$ symmetry. A well-known example of this duality is the 1D transverse field Ising model~\cite{Sachdev2011}, where the elementary excitation in the broken symmetry phase is the creation of a domain wall, exhibiting the $Z_2$ symmetry.

In quantum many-body spin systems, the appearance of a completely factorized ground state in the presence of strong spin-spin interactions has generated much attention~\cite{vidal2004entanglement, giampaolo2008theory,giampaolo2009separability, roscilde2005entanglement, rezai2010factorized, karpat2014quantum, cui2008multiparticle}. This so-called ground-state factorization occurs in the broken-symmetry phase of interacting spin systems. Despite pioneering studies showing its occurrence in various types of spin models, as well as the development of a quantum information theoretic diagnostic tool to detect the factorizing point~\cite{giampaolo2013universal, ccakmak2015factorization, su2022long, cheng2015criticality}, the physical origin of the emergence of ground-state factorization has remained elusive. Moreover, the studies on ground-state factorization and its diagnostic tools have thus far been limited to spin systems. Therefore, it is unclear whether a generic quantum many-body system involving continuous variables such as harmonic oscillators admits ground-state factorization or not.

In this Letter, we present a theory in which a continuous symmetry may emerge in the broken $Z_2$ symmetry phase, and we identify it as a physical origin of the ground-state factorization. Using a generic asymmetric Landau potential in phase space, which provides an exact description of phase transitions in fully-connected systems, we establish that a continuous symmetry could emerge from the breaking of $Z_2$ symmetry. Based on this insight, we discover that the Dicke model~\cite{hamner2014dicke, dicke1954coherence, baumann2010dicke, hioe1973phase, soldati2021multipartite,vidal2006finite,barthel2006entanglement,vidal2007entanglement,baksic2014controlling, Safavi-Naini.2018,zhang2018dicke,Bakemeier:2012ja,Baumann:2011io}, a paradigmatic model undergoing a superradiant phase transition, exhibits lines of emergent continuous symmetry in the presence of anisotropic interaction. Our analysis shows that continuous symmetry emerges as the order parameter renormalizes either the particle number preserving or non-preserving coupling strength to be zero. When the renormalized particle number non-conserving interaction vanishes, the emergent conservation law for the total particle number leads to a completely factorized ground state. We note that this is the first example of ground-state factorization going beyond spin systems, and it is also a striking feature that has gone unnoticed in previous studies of the paradigmatic quantum optical model. Furthermore, we establish that the ground-state factorization predicted for the anisotropic Lipkin-Meshkov-Glick (LMG) model~\cite{vidal2004entanglement, giampaolo2008theory,giampaolo2009separability} is also a consequence of the emergent continuous symmetry. Finally, we propose identifying the emergent continuous symmetry as a diagnostic tool for the ground-state factorization point. In many-body systems for which it is challenging to obtain the exact ground-state solution, identifying the emergent continuous symmetry point may still be possible, at which the model becomes integrable. We demonstrate this possibility by showing the ground-state factorization of the Dicke lattice model in arbitrary dimensions and lattice geometries.

\begin{figure}[h]
    \centering
    \includegraphics[width = 0.5\textwidth]{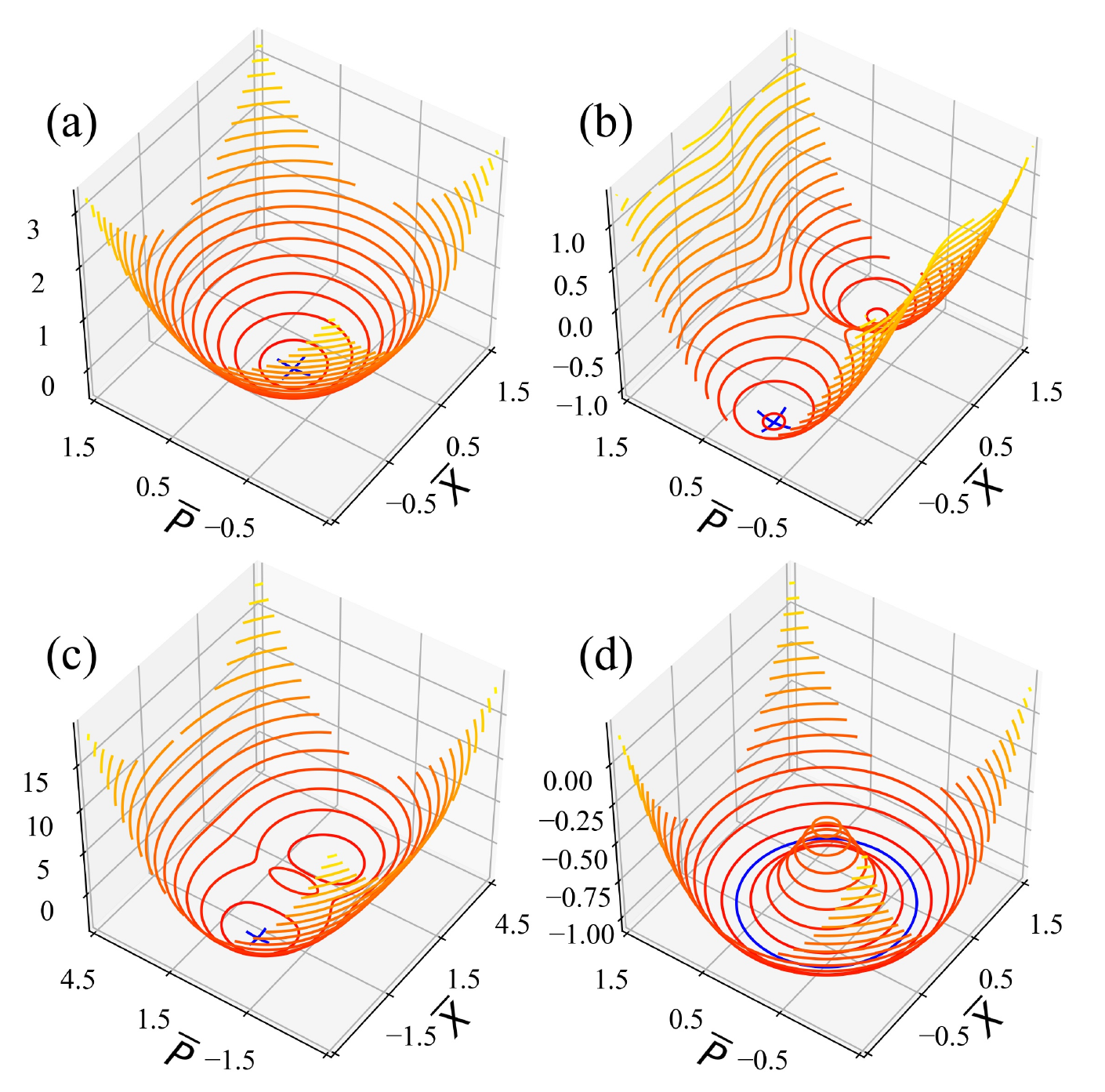}
    \caption{The mean-field energy landscape as a function of the real and imaginary  part of a dimensionless complex order parameter $\bar\alpha=\bar X+i \bar P$. (a) The normal phase with $\bar{X} = \bar{P} = 0$ as the sole minimum. (b) A specific instance and (c) a generic instance of the broken discrete symmetry phase, where the principal curvatures at the local minima (blue) are equal and different, respectively. (d) The broken continuous symmetry leads to a circle of degenerate minima, leading to the Goldstone mode.}
    \label{fig:demo}
\end{figure}

{\it Landau theory of emergent continuous symmetry.---} In the spirit of the Landau theory of second-order phase transitions, we consider a mean-field (MF) energy potential whose minimum determines a dimensionless order parameter $\bar\alpha=\bar X +i\bar P$,
\begin{equation}
\label{MF_energy}
    \bar{E}_\textrm{MF} (\bar X, \bar P) = \bar{X}^2 + \bar{P}^2 - \frac{1}{2}\sqrt{1 + (2g_+\bar{X})^2 + (2g_-\bar{P})^2}.
\end{equation}
In physical systems, $\bar\alpha$ may correspond to the spontaneous coherence of the cavity field or the spontaneous polarization of collective spins. For $g_+=g_-$, the MF energy has a continuous $U(1)$ symmetry since it becomes invariant under $(\bar X,\bar P)\rightarrow (\cos\phi \bar X-\sin\phi \bar P, \sin \phi \bar X+\cos\phi \bar P)$ for any $\phi\in[0,2\pi)$. When $g_+=g_->1$, the continuous symmetry breaking leads to the gapless Goldstone mode since rotation around the minimum of the Mexican hat potential incurs no energy cost [Fig.~\ref{fig:demo} (d)]. For $g_+\neq g_-$, the symmetry of the MF energy is reduced to a $Z_2$ symmetry (invariant under $\bar X(\bar P)\rightarrow-\bar X(\bar P)$), which is spontaneously broken for $\max(|g_+|,|g_-|) > 1$. The order parameters, determined by the positions of the new minima, are $\{\bar X_0,\bar P_0\}=\{\pm \frac{1}{2g_+}\sqrt{g_+^4-1},0\}$ for $\max(|g_+|,|g_-|)=|g_+|>1$ and $\{0,\pm \frac{1}{2g_-}\sqrt{g_-^4-1}\}$ for $\max(|g_+|,|g_-|)=|g_-|>1$. Without a loss of generality, let us focus on $\max(|g_+|,|g_-|)=|g_+|>1$ [See Fig.~\ref{fig:demo}(b,c)]. Expanding the MF energy around the new minima $\bar X=\bar X_0+\bar x$ and $\bar P=\bar P_0+\bar p$ up to the second order of $\bar x$ and $\bar p$, we obtain the energy potential for the fluctuations.
\begin{equation}
\label{MF_fluc}
    \bar{E}_\textrm{fluc} (\bar x, \bar p) = (1-\frac{1}{g_+^4})\bar x^2+(1-\frac{g_-^2}{g_+^2})\bar p^2.
\end{equation}
The excitation in the broken symmetry phase, therefore, has the $Z_2$ symmetry for most of the parameters since Eq.~(\ref{MF_fluc}) is invariant under $\bar x (\bar p)\rightarrow-\bar x (\bar p)$, which is typically expected for the $Z_2$ symmetry breaking [Fig.~\ref{fig:demo} (c)]~\cite{ribeiro2007thermodynamical,ribeiro2008exact}. Interestingly, however, for $|g_+g_-|=1$, the curvatures along $\bar x$ and $\bar p$ become identical [Fig.~\ref{fig:demo} (b)]; that is, the potential for fluctuation acquires a higher symmetry (continuous symmetry) than the original MF potential, since Eq.~(\ref{MF_fluc}) becomes invariant under $(\bar x,\bar p)\rightarrow (\cos\phi \bar x-\sin\phi \bar p, \sin \phi \bar x+\cos\phi \bar p)$ for any $\phi\in[0,2\pi)$. Our analysis reveals an interesting possibility in which the excitation in the broken symmetry phase of discrete $Z_2$ symmetry may acquire a continuous symmetry. 

Below, we show that the MF energy we introduced in Eq.~\eqref{MF_energy} is, in fact, the mean-field energy of a wide range of fully-connected models with anisotropic interaction, where the strengths of the particle number preserving and non-preserving interactions are not equal; these include, but are not limited to, the anisotropic Dicke model and its lattice generalization, as well as the anisotropic LMG model. Since the mean-field approximation for such models becomes exact in the corresponding thermodynamic limit, we predict that the emergent continuous symmetry should appear in the broken symmetry phase. 

{\it Anisotropic Dicke model.---} Motivated by our prediction based on the mean-field energy, let us first consider the anisotropic Dicke model, where a harmonic oscillator is coupled to $N$ spin-1/2 particles.
\begin{equation}
\label{Dicke}
	H^\mathrm{Dicke} = \omega_0a^{\dagger}a + \Omega J_z + \frac{2\lambda_+}{\sqrt{N_a}}(a^{\dagger}+a)J_x - i\frac{2\lambda_-}{\sqrt{N_a}}(a^{\dagger} - a)J_y.
\end{equation}
Here, $a^\dagger$ and $a$ are the creation/annihilation operators of the oscillator, and $J_{x,y,z}$ are the collective spin operators for $N_a$ spins. The superradiant phase transition of $H^\mathrm{Dicke}$ has been intensively studied in both the $N_a\rightarrow\infty$~\cite{emary2003quantum, lambert2004entanglement,baksic2014controlling}. For $\lambda_+=\lambda_-$, the continuous symmetry-breaking phase transition leads to the emergence of the Goldstone mode~\cite{baksic2014controlling}. For $\lambda_+\neq\lambda_-$, $H^\mathrm{Dicke}$ has the $Z_2$ symmetry-breaking phase transition~\cite{emary2003quantum, lambert2004entanglement,baksic2014controlling}. We revisit the phase diagram of $H^\mathrm{Dicke}$ and uncover previously unnoticed striking features, namely the emergent continuous symmetry and the ground-state factorization. 

\begin{figure}[h]
    \centering 
    \includegraphics[width = 0.4\textwidth]{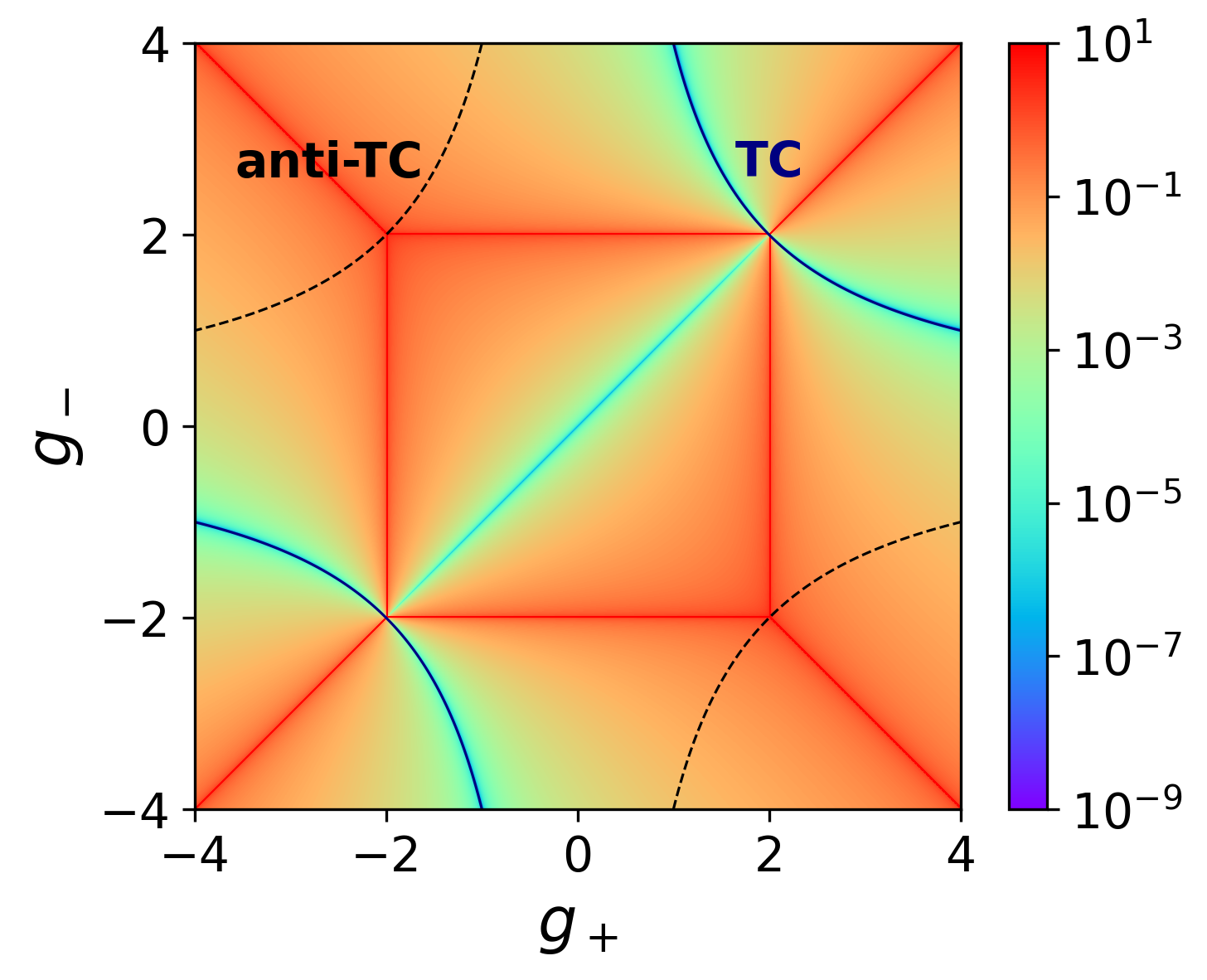}
    \caption{The ground-state entanglement diagram for the anisotropic Dicke model in $g_{+}$-$g_{-}$ plane. In addition to the well-known factorized ground state along the line of $g_+=g_-$ in the normal phase, the ground-state factorization occurs also in the $Z_2$-symmetry broken superradiant phase along the emergent Tavis-Cummings line ($g_+g_-=1$) where the conversation law for the total number of particles, absent in the bare Hamiltonian, appears. The emergent anti-TC line ($g_+g_-=-1$, dashed line), on the other hand, does not feature the ground-state factorization.}
    \label{fig: Dicke density}
\end{figure}

To this end, we obtain the mean-field energy of $H^\mathrm{Dicke}$ by replacing $a$ with $\alpha = \sqrt{\frac{N\Omega}{\omega_0}}(\bar{X} + i\bar{P}) \in \mathbb{C}$ and $(J_x,J_y,J_z)\rightarrow\frac{N}{2}(\sin\theta\cos\varphi, \sin\theta\sin\varphi, \cos\theta)$ with $\theta \in[0,\pi],\varphi\in[0,2\pi)$ and $H^\mathrm{Dicke} \rightarrow N\Omega\bar{E}^\mathrm{Dicke}$, which leads to
\begin{equation}\label{dicke_mf}
	\bar{E}^\mathrm{Dicke} = \bar{X}^2 + \bar{P}^2 + g_+\bar{X}\sin\theta\cos\varphi - g_-\bar{P}\sin\theta\sin\varphi,
\end{equation}
up to a constant $\frac{1}{2}\cos\theta$, where $g_\pm \equiv 2\lambda_\pm/\sqrt{\omega_0\Omega}$. When $\max(|g_+|,|g_-|) < 1$, the energy minimum is at $\alpha_\textrm{np} = 0$, $\theta_\textrm{np} = \pi$ for any $\varphi$, i.e., the normal phase. Otherwise, the spins acquire a spontaneous polarization given by $\varphi_\textrm{sp} = - \arctan\left(\frac{g_-\bar{P}}{g_+\bar{X}}\right)$ and $\theta_\textrm{sp} = \pi - \arctan\left(2\sqrt{(g_+\bar{X})^2 + (g_-\bar{P})^2}\right)$. By plugging $\theta$ and $\varphi$ solutions into Eq.~(\ref{dicke_mf}), $\bar{E}^\mathrm{Dicke}$ becomes exactly the same as Eq.~\ref{MF_energy}, for which an emergent continuous symmetry at $|g_+g_-|=1$ is predicted. To investigate its consequences on the quantum properties, we perform a unitary transformation $\tilde H^\mathrm{Dicke} = U^\dagger H^\mathrm{Dicke}U$ with $U=\mathrm{e}^{-i\varphi J_z}\mathrm{e}^{-i\theta J_{y}}\mathrm{e}^{\alpha a^{\dagger}-\alpha^*a}$ using MF solutions $\phi$, $\theta$, and $\alpha$. Here, we focus on $|g_+|>|g_-|$ case and treat $|g_+|<|g_-|$ in the Supplementary Material. In the superradiant phase, the transformed Hamiltonian, followed by the Holstein-Primakoff transformation $J_+ = \sqrt{N - b^\dagger b}b $ and $J_z = \frac{N}{2}-b^{\dagger}b$, reads: 
\begin{align}\label{dicke_sp}
H_\textrm{sp}^\mathrm{Dicke} &= \omega_0a^{\dagger}a +  \tilde\Omega b^\dagger b\\\nonumber
& \quad+(\widetilde{\lambda}_++\lambda_-)(a b^\dagger+a^\dagger b)+(\widetilde{\lambda}_+-\lambda_-)(a b+a^\dagger b^\dagger)
\end{align}
where  $\widetilde{\Omega} = g_+^2\Omega$ and $\widetilde{\lambda}_+ = \frac{\sqrt{\omega_0\Omega}}{2g_+}$. At $|g_+g_-|=1$, where we predicted the emergence of continuous symmetry based on Landau's theory, the renormalized coupling becomes $\lambda_-=\pm\tilde\lambda_+$. For $g_+g_-=1$, i.e., $\tilde\lambda_+=\lambda_-$, the counter-rotating term $a b +a^\dagger b^\dagger$ vanishes, and $H_\textrm{sp}^\mathrm{Dicke}$ reduces to the same form as the Tavis-Cummings (TC) Hamiltonian in the SP phase. Therefore, $N_\textrm{sum}=a^\dagger a+b^\dagger b$ becomes a conserved quantity that leads to the emergent continuous symmetry (invariant under $a\rightarrow a e^{i\theta}$ and $b\rightarrow b e^{i\theta}$). Moreover, the conservation of $N_\textrm{sum}$ makes the ground-state a simple product state of vacuum $|00\rangle_\textrm{ab}$, thus leading to ground-state factorization. For $g_+g_-=-1$, i.e., $\tilde\lambda_+=-\lambda_-$, on the other hand, the rotating term $a b^\dagger+a^\dagger b$ vanishes, leading to the anti-TC Hamiltonian. In this case, $N_\textrm{diff}=a^\dagger a-b^\dagger b$ is the conserved quantity, which also leads to the emergent continuous symmetry (invariant under $a\rightarrow a e^{i\theta}$ and $b\rightarrow b e^{-i\theta}$). The crucial difference is that the conservation of $N_\textrm{diff}$ does not make the ground-state a factorized state. In Fig.~\ref{fig: Dicke density}, we present the ground-state entanglement measured by the von Neumann entropy, $S_v(\rho_g)=-\rho_g \mathrm{Tr}[\log_2\rho_g]$. The ground-state factorization along the emergent TC line ($g_+g_-=1$) in the $Z_2$-symmetry broken phases is clearly demonstrated. Our analysis, therefore, establishes that ground-state factorization in the anisotropic Dicke model occurs precisely when a continuous symmetry emerges, owing to the conservation of the total number of excitations. In contrast, the emergent continuous symmetry associated with the conservation of the difference in excitation numbers does not lead to ground-state factorization, as evidenced by the nonvanishing entanglement observed along the emergent anti-TC line.

We note that the effective Hamiltonian in the SP phase can also be obtained by performing the Holstein-Primakoff transformation first, followed by displacing the boson operator, as done in Ref.~\cite{baksic2014controlling}. In this case, one must apply an additional local squeezing transformation to obtain $H_\textrm{sp}^\textrm{Dicke}$ given in Eq.~\eqref{dicke_sp}, where the emergent continuous symmetry becomes apparent~\cite{supp}.

{\it Ground-state factorization in the Dicke lattice model.---} Having demonstrated the emergent continuous symmetry and its associated conservation law as the condition for ground-state factorization in the single-mode anisotropic Dicke model, we now turn to the Dicke lattice model, described by 
 \begin{equation}\label{DickeLattice}
     H^\mathrm{DL} = \sum_{j}H^\mathrm{Dicke}_{j} + \frac{J}{2}\sum_{\langle j,k\rangle}\left(a_{j}^\dagger a_{k} + a_{j}a_{k}^\dagger\right).
 \end{equation}
Here, $H^\mathrm{Dicke}_{j}=\omega_0a_j^{\dagger}a_j + \Omega J^j_z + \frac{2\lambda_+}{\sqrt{N}}(a_j^{\dagger}+a_j)J^j_x - i\frac{2\lambda_-}{\sqrt{N}}(a_j^{\dagger} - a_j)J^j_y$ is the on-site Dicke Hamiltonian, and the second term represents the nearest-neighbor photon-hopping interaction. The model is defined on a lattice of arbitrary spatial dimensions with periodic boundary conditions. We choose $J<0$ in order to avoid the frustrated superradiant phase transition~\cite{zhao2022frustrated}, which can lead to extensive ground-state degeneracy and intricate entanglement among the degenerate configurations.

The mean-field energy is obtained by using local mean-values $\alpha_j$ and $\{\theta_j,\varphi_j\}$, for the bosonic mode and the spins at site $j$, respectively. Since the $j$-th spins are only coupled to their on-site $j$-th bosonic mode, the spin variables can be eliminated exactly as in the single Dicke case, yielding a boson-only mean-field energy for the Dicke lattice model,
\begin{align}
\bar{E}^\mathrm{DL} = \sum_{j}\bar{E}_{j} + \frac{J}{\omega_0}\sum_{\langle j,k\rangle}\left(\bar{X}_{j}\bar{X}_{k} + \bar{P}_{j}\bar{P}_{k}\right)
\end{align}
with 
\begin{align}
\bar{E}_{j} = \bar{X}_{j}^2 + \bar{P}_{j}^2 - \frac{1}{2}\sqrt{1 + (2g_+\bar{X}_{j})^2 + (2g_-\bar{P}_{j})^2}.
\end{align}
Using Jensen’s inequality for the concave on-site square-root term and the Cauchy–Schwarz inequality for the hopping term, we show that any spatial modulation increases the total energy, and the lower bound is saturated only by the uniform configuration $\bar{X}_j=\bar{X}$ and $\bar{P}_j=\bar{P}$~\cite{supp}. Consequently, the mean-field energy of the Dicke lattice also reduces to the Landau potential in Eq.~\eqref{MF_energy} with renormalized coefficients for quadratic terms,
\begin{align}
\frac{\bar{E}^\mathrm{DL}}{N}
=(1+\frac{Jz}{\omega_0})(\bar{X}^2+\bar{P}^2)
-\tfrac{1}{2}\sqrt{1+(2g_+\bar{X})^2+(2g_-\bar{P})^2}.
\end{align}
where $z$ is the coordination number and $N$ is the lattice size. Therefore, the mean-field phase diagram becomes identical to that of the single Dicke model, except for the shifted critical point, $g_c = \sqrt{1+\frac{Jz}{\omega_0}}$.
Moreover, we identify that an emergent continuous symmetry appears when 
\begin{equation}
    |g_+ g_-| = g_c^2=1+\frac{Jz}{\omega_0},
\end{equation}
which signals the existence of a ground-state factorization in the many-body ground state of the Dicke lattice model. Indeed, when \( g_+ g_- = g_c^2 \), the effective Hamiltonian in the superradiant (SP) phase—obtained by displacing the bosonic modes and rotating the spins, followed by the Holstein–Primakoff transformation—contains only particle-conserving terms \(a_j^\dagger b_j + a_j b_j^\dagger\), in the on-site spin–boson interaction, where \(b_j\) denotes the Holstein–Primakoff boson for the local collective spins. Since the photon-hopping interaction preserves the total excitation number, this quantity becomes an emergent conserved quantity, and the ground state is fully factorized as a product of local vacuum states.
\begin{equation}
\ket{\Psi_g}
= \bigotimes_{j}\left(
  \ket{0}_j \otimes
  \ket{N_a/2, -N_a/2}_{j}
\right).
\end{equation}
Our analysis thus shows that the ground-state factorization at \( g_+ g_- = g_c^2 \) is a universal feature of the Dicke lattice model with negative photon-hopping energy, independent of the lattice geometry or dimensionality. Moreover, it successfully demonstrates that identifying the emergent continuous-symmetry points provides a diagnostic tool for locating the ground-state factorization point in quantum many-body systems.

\begin{figure}[h]
    \centering
    \includegraphics[width = 0.5\textwidth]{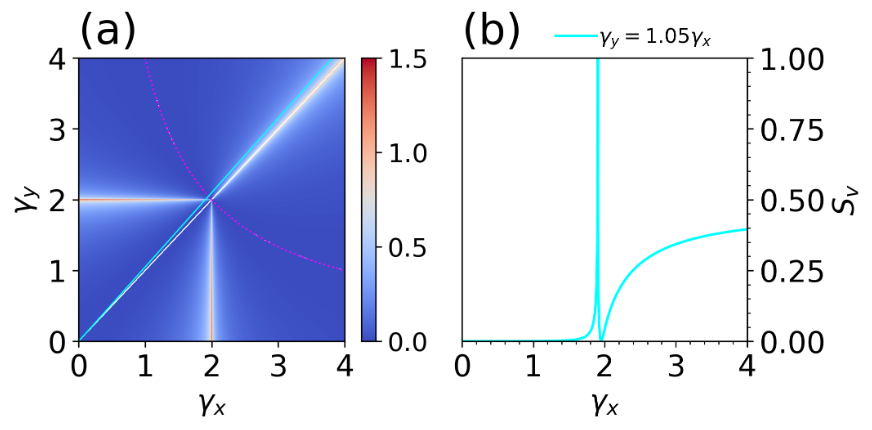}
    \caption{(a) The ground-state entanglement diagram for the LMG model. The emergent continuous symmetry appears along the dashed magenta line. (b) The entanglement along the line (cyan) $\gamma_y = 1.05\gamma_x$ indicated in (a) shows the ground-state factorization as it crosses the emergent continuous symmetry line.}
    \label{fig:LMG}
\end{figure}

{\it LMG model.---} The LMG model serves as a paradigmatic example of a fully connected quantum spin system, for which the mean-field analysis becomes exact in the thermodynamic limit~\cite{ribeiro2008exact,ribeiro2007thermodynamical}, and the ground-state entanglement and factorization have been extensively studied~\cite{vidal2004entanglement,dusuel2004finite,dusuel2005continuous,barthel2006entanglement,latorre2005entanglement,giampaolo2008theory,giampaolo2009separability}. We now revisit the ground-state factorization of the LMG model from the perspective of Landau theory of emergent continuous symmetry, thereby providing a unified understanding with the qubit-oscillator systems discussed above. Let us consider the LMG Hamiltonian,
\begin{equation}\label{LMG}
	H^\mathrm{LMG} = -hJ_z - \frac{\gamma_x}{2j}J_x^2 - \frac{\gamma_y}{2j}J_y^2,
\end{equation}
where $h>0$ denotes the transverse magnetic field strength, $\gamma_{x,y}$ is the spin-spin interaction strength, and $J_{x,y,z}$ is the collective spin operator for $N=2j$ spin-$1/2$ particles. By using the spin-coherent state, $
\ket{\theta,\varphi} = e^{-i\varphi J_z} e^{-i\theta J_y} \ket{j}$
, where $J_z\ket{j}=j\ket{j}$ and $\{\theta,\varphi\}$ are polar coordinates, the collective spin expectation values read
 $(J_x, J_y, J_z)=j(\sin\theta\cos\varphi,\sin\theta\sin\varphi,\cos\theta)$. By replacing the operators with these mean-values, we obtain the mean-field energy $\bar E^\textrm{LMG}=E^\textrm{LMG}/(jh)=
    E^\mathrm{LMG} = -\cos\theta - \frac{1}{2h}(\gamma_x\cos^2\varphi + \gamma_y\sin^2\varphi)\sin^2\theta$. Using $X = \sin\theta\cos\varphi$ and $Y=\sin\theta\sin\varphi$,
\begin{equation}
    \bar{E}^\mathrm{LMG} = -\frac{\gamma_x}{2h}X^2 - \frac{\gamma_y}{2h}Y^2 - \sqrt{1 - X^2 - Y^2}.
\end{equation}
If $\gamma_{x,y} < h$, the normal phase $X = Y = 0$ minimizes energy. The $Z_2$-symmetry broken phase emerges when  $\max(\gamma_x,\gamma_y)>h$. Focusing on $\gamma_x\geq\gamma_y$ case, the mean-field solutions are $X_0^2 = 1 - \left(\frac{h}{\gamma_x}\right)^2$, $Y_0 = 0$, or equivalently, $\cos\theta_0 = \frac{h}{\gamma_x},\ \varphi_0 = 0$. Considering small fluctuations around the mean-field solution, $x=X-X_0$ and $y=Y-Y_0$, the energy potential for fluctuations reads $\bar E_\mathrm{fluc}(x,p)=\frac{\kappa_x}{2}x^2+\frac{\kappa_y}{2} y^2$, where the quadratic coefficients are 
\begin{equation}
\kappa_x=\gamma_x(\frac{\gamma_x^2}{h^2}-1),\quad \kappa_y =\gamma_x-\gamma_y.
\end{equation}
Note that the rotational invariance of the fluctuation energy on a Bloch sphere around a mean-field solution exists when it is proportional to the line element:
\begin{equation}
ds^2=d\theta^2+\sin^2\theta_0 d\varphi^2=\frac{x^2}{\cos\theta^2_0} +{y^2}
\end{equation}
where we used $x=\cos\theta_0 d\theta$ and $y=\sin\theta_0 d\varphi$. Therefore, the emergent continuous symmetry appears when
\begin{equation}
\kappa_x\cos^2\theta_0=\kappa_y\rightarrow \gamma_x\gamma_y=h^2.
\end{equation}
The latter condition is precisely the ground-state factorization point of the LMG model~\cite{vidal2004entanglement,giampaolo2008theory}. To identify the associated conserved quantity, we rotate the spin by a unitary operator $U = \mathrm{e}^{-i\varphi_0 J_z}\mathrm{e}^{-i\theta_0 J_{y}}$, which leads to 
\begin{equation}
    \tilde H^\mathrm{LMG} = -\frac{h^2}{\gamma_x}J_z - \frac{1}{2j}\left(\gamma_x - \frac{h^2}{\gamma_x}\right)J_z^2 - \frac{h^2/\gamma_x}{2j}J_x^2 - \frac{\gamma_y}{2j}J_y^2.
\end{equation}
Following Ref.~\cite{vidal2007entanglement}, the collective spins can be divided into two equal parts, followed by the Holstein-Primakoff mapping for each $j$ spin to derive the quadratic Hamiltonian of two coupled Holstein-Primakoff bosons $b_1$ and $b_2$. The bipartite ground-state entanglement measured by the von Neumann entropy ($S_v$) is presented in Fig.~\ref{fig:LMG}. The interaction term between the $b_1$ and $b_2$ modes is proportional to 
\begin{equation}
H_{12}\propto\left(\frac{h^2}{\gamma_x}+\gamma_y\right)(b_1b_2^\dagger +b_2 b_1^\dagger)+\left(\frac{h^2}{\gamma_x}-\gamma_y\right)(b_1b_2 +b_1^\dagger b_2^\dagger).
\end{equation}
The second term, a particle non-conserving term, vanishes for $h^2=\gamma_x\gamma_y$, confirming that the emergent continuous symmetry in the mean-field energy is due to the emergent conservation law of $N_\textrm{sum}=b_1^\dagger b_1+b_2^\dagger b_2$.

{\it Conclusion.---} We have established a generic mechanism by which an emergent continuous symmetry arises in the broken phase of a system with only discrete symmetry. This symmetry emerges from the renormalization of anisotropic long-range couplings by the order parameter, which generates an effective conservation law absent in the bare Hamiltonian and leads to ground-state factorization. Applying this framework, we uncovered the ground-state factorization of the anisotropic Dicke model—the first example of factorization in a system with continuous variables—and identified the physical origin of factorization in the LMG model. Furthermore, we demonstrated that the emergence of continuous symmetry serves as a powerful diagnostic for locating the ground-state factorization point in the Dicke lattice models. It would be interesting to explore whether our framework can be extended to the ground-state factorization of quantum many-body systems with short-range interactions.

{\it Acknowledgments.---}This work was supported  by the Innovation Program for Quantum Science and Technology 2021ZD0301602 and the Summer Research Scholar Program from Duke Kunshan University.

\bibliography{bib.bib}
\end{document}